\documentstyle[12pt]{article}
\begin{document}
\def\om{\omega}
\def\omt{\tilde{\omega}}
\def\ti{\tilde}
\def\o{\Omega}
\def\t{T^*M}
\def\vt{\tilde{v}}
\def\ot{\tilde{\Omega}}
\def\otwo{\omt \wedge \om}
\def\owot{\om \wedge \omt}
\def\w{\wedge}
\def\mt{\tilde{M}}

\def\om{\omega}
\def\omt{\tilde{\omega}}
\def\ss{\subset}

\def\om{\omega}
\def\omt{\tilde{\omega}}
\def\ti{\tilde}
\def\o{\Omega}
\def\t{T^*M}
\def\vt{\tilde{v}}
\def\ot{\tilde{\Omega}}
\def\otwo{\omt \wedge \om}
\def\owot{\om \wedge \omt}
\def\w{\wedge}
\def\mt{\tilde{M}}

\def\om{\omega}
\def\omt{\tilde{\omega}}
\def\ss{\subset}
\def\tpm{T_{P} ^* M}
\def\al{\alpha}
\def\alt{\tilde{\alpha}}
\def\la{\langle}
\def\ra{\rangle}
\def\inop{{\int}^{P}_{P_{0}}{\om}}
\def\th{\theta}
\def\tht{\tilde{\theta}}
\def\inox{{\int}^{X}{\om}}
\def\inotx{{\int}^{X}{\omt}}
\def\st{\tilde{S}}
\def\ls{\lambda_{\sigma}}
\def\p{{\bf{p}}}
\def\pb{{\p}_{b}(t,u)}
\def\pbm{{\p}_{b}}
\def\d{\partial}
\def\d+{\partial_+}
\def\d-{\partial_-}
\def\pat{\partial_{\tau}}
\def\pas{\partial_{\sigma}}
\def\dpm{\partial_{\pm}}
\def\l2{\Lambda^2}
\def\be{\begin{equation}}
\def\ee{\end{equation}}
\def\bea{\begin{eqnarray}}
\def\eea{\end{eqnarray}}
\def\ej{{\bf E}}
\def\ed{{\bf E}^\perp}
\def\si{\sigma}
\def\cg{{\cal G}}
\def\cgt{\ti{\cal G}}
\def\cd{{\cal D}}
\def\ce{{\cal E}}
\def\cep{\ce^{\perp}}
\def\cf{{\cal F}}
\def\cfp{\cf^{\perp}}
\def\bz{\bar{z}}
\def\e{\varepsilon}
\def\b{\beta}
\begin{titlepage}
\begin{flushright}
{}~
CERN-TH/96-142\\
hep-th/9605212
\end{flushright}

\vspace{4cm}
\begin{center}
{\Large \bf  Non-Abelian Momentum-Winding Exchange}\\
[50pt]{\small
{\bf C. Klim\v{c}\'{\i}k}\\
Theory Division, CERN,\\ 
CH-1211 Geneva 23, Switzerland \\[3pt] and \\[3pt] 
{\bf P. \v Severa }\\
Department of Theoretical Physics, Charles University, \\
V Hole\v sovi\v ck\'ach 2, CZ-18000 Praha,
Czech Republic\\[30pt] }

\begin{abstract}
\noindent
A non-Abelian analogue of the Abelian $T$-duality momentum-winding exchange
is described. The non-Abelian $T$-duality relates $\sigma$-models living
on the cosets of a Drinfeld double with respect to its isotropic subgroups.
The role of the Abelian momentum-winding lattice is in general played by the
fundamental group of the Drinfeld double.

\end{abstract}
\end{center}
\vskip 2.5cm
\noindent CERN-TH/96-142\\
May 1996

\end{titlepage}

 1. The Poisson-Lie (PL) $T$-duality \cite{KS2} is a generalization of the 
traditional non-Abelian $T$-duality \cite{OQ}--\cite{GR} and it  proved
to enjoy 
\cite{KS2},
\cite{T}--\cite{KS4}  most of the
structural features of the traditional Abelian $T$-duality 
\cite{SS}--\cite{Busch}.

The purpose of this note is to settle the global issues of the PL 
$T$-duality for closed strings.
From the space-time point of view, we shall 
identify the targets of the mutually 
dual $\sigma$-models
with the cosets $D/G$ and $D/\ti G$, where $D$ denotes the Drinfeld double
and $G$ and $\ti G$ two its isotropic subgroups.  In the special
case when the decomposition $D=G\ti G=\ti GG$ holds globally the 
corresponding
cosets turn out to be the group manifolds $\ti G$ and $G$ respectively 
\cite{KS2}.
Then we shall describe the momentum and the winding states from the point
of view of both targets $D/G$ and $D/\ti G$ and show how the PL $T$-duality
interchange them.

 2. 
For the description of the Poisson-Lie $T$-duality we need the
crucial concept
of the Drinfeld double,  which is simply  a  Lie group $D$ such that
its Lie algebra $\cd$ (viewed as a vector space) 
 can be decomposed as the direct sum  of two  subalgebras $\cg$ and $\cgt$ 
maximally isotropic with 
respect to a non-degenerate invariant bilinear form on $\cd$ \cite{D}.
It is often convenient to identify the dual linear space to $\cg$ ($\cgt$)
with $\cgt$($\cg$) via this bilinear form.

Consider the right coset $LD/D$ where $LD$ denotes the loop group
of the Drinfeld double. There is a natural symplectic
 two-form $\Omega$ on $LD/D$ \cite{KS3} given
as the exterior derivative of a  polarization one-form $\alpha$.
 The latter
is most naturally defined in terms of its integral along an arbitrary
curve $\gamma$ in the phase space, parametrized by a parameter $\tau$.
From the point of view of the Drinfeld double, this curve is a surface 
$l(\tau,\sigma)\in D$ with
the topology of a cylinder, embedded in the double.  We define
\be \int_{\gamma} \alpha ={1\over 8\pi}\int \la \pas l~l^{-1},
\pat l~l^{-1}\ra+{1\over 48\pi} \int d^{-1}\la dl~l^{-1},[dl~l^{-1},
dl~l^{-1}]\ra .\ee
Here $\la .,.\ra $ denotes the non-degenerate invariant bilinear form
on the Lie algebra ${\cal D}$  of the double; in  the second term
 on the r.h.s., we
recognize
the two-form
potential
of the WZW three-form on the double. Note that this definition of $\alpha$ is
ambiguous since the choice of 
the inverse exterior derivative  $d^{-1}$ is too. However, this ambiguity 
disappears
when the exterior derivative of the one-form $\alpha$ is taken. In other words,
the symplectic form $\Omega$ is well defined.

In this note, we  study a dynamical system on the phase space $P\equiv LD/D$
given by the action\footnote{This action 
$S$ has also  a little gauge invariance corresponding to the right 
multiplication
of $l(\si,\tau)$ by an arbitrary function $l(\tau)\in D$. This small gauge
symmetry corresponds to the factorization $LD/D$.}
$$ S[l(\tau,\si)]=\int \alpha - \int H~d\tau $$
\be ={1\over 8\pi}\int d\si d\tau \biggl\{\la \pas l~l^{-1},\pat l~l^{-1}\ra+
{1\over 6}d^{-1}\la dl~l^{-1},[dl~l^{-1},
dl~l^{-1}]\ra -\la \pas l l^{-1},R\pas l l^{-1}\ra \biggl\}.\ee
Here $R$ is a linear idempotent selfadjoint map from the Lie Algebra
$\cd$ of the double into itself. It has two equally degenerated
eigenvalues $+1$ and $-1$ and we choose in $\cd$ some orthonormal basis
$R_+^a$ and $R_-^a$ of the corresponding eigenvectors\footnote{The condition
of the positive or negative
definiteness of the form $\la .,.\ra$ on the subspaces $Span(R_{\pm}^a)$
can be easily released; however, the form should be still 
non-degenerate there!}:
\be \la R_{\pm}^a,R_{\pm}^b\ra= \pm\delta^{ab},\qquad \la R_+^a,R_-^b\ra=0.\ee
 Note that
\be R=\vert R_+^a\ra\la R_+^a \vert + \vert R_-^a\ra\la R_-^a\vert,\ee
\be Identity= \vert R_+^a\ra\la R_+^a \vert -\vert R_-^a\ra\la R_-^a\vert.\ee
We may also remark that the first two terms of the first-order action 
$S$ give together the standard WZNW action on the double if we
interpret $\tau$ and $\sigma$ as the `light-cone' variables.
This means that we can conveniently use the Polyakov-Wiegmann formula
\cite{PW}.

Let $G$ be an $n$-dimensional  subgroup of 
$2n$-dimensional Drinfeld double $D$ such that the Lie algebra $\cg$ od $G$ 
is isotropic (i.e. $\la \cg, \cg \ra=0$). Consider then the  right coset
$D/G$ and parametrize it by the elements $f$ of $D$ \footnote{If there
exists no global section of this fibration, we can choose several
local sections covering the whole base space $D/G$.}.
With this parametrization of $D/G$ we may parametrize the surface
$l(\tau,\sigma)$ in the double as follows
\be l(\tau,\sigma)= f(\tau,\sigma)g(\tau,\sigma),\quad  g\in G.\ee
The action $S$ then becomes 
$$ S(f,\Lambda\equiv \pas g g^{-1})={1\over 2}I(f) -{1\over 2\pi}
\int d\xi^+ d\xi^- \biggl\{\la \Lambda -
{1\over 2} f^{-1}\d- f , \Lambda -{1\over 2}f^{-1}\d- f \ra$$
\be +\la f\Lambda f^{-1} +\pas f f^{-1}, R_-^a\ra\la R_-^a , f\Lambda f^{-1}
+\pas f f^{-1}\ra\biggl\},\ee
Here
\be I(f)\equiv {1\over 4\pi}\int d\xi^+ d\xi^-\la \partial_+ f~f^{-1}, 
\partial_- f~f^{-1}\ra +{1\over 24\pi} 
\int d^{-1}\la df~f^{-1},[df~f^{-1},
df~f^{-1}]\ra \ee
is the standard WZNW action in the proper light-cone variables, defined
by 
\be \xi^{\pm}\equiv{1\over 2}(\tau \pm \sigma).\ee 

It is now easy to solve $\Lambda$ in terms of $f$:
\be \Lambda_b T^b ={1\over 2}\la \partial_- f~f^{-1}, 
R_+^a\ra (M_+^{-1})_{ab}T^b-
{1\over 2}\la \partial_+ f~f^{-1},R_-^a\ra (M_-^{-1})_{ab}T^b,\ee
where $T^b$ is some basis of $\cg$ and
\be M_{\pm}^{ab}\equiv \la fT^a f^{-1},R_{\pm}^b\ra.\ee
Inserting this expression back into (7), we obtain the following 
second-order action of a $\sigma$-model living on the coset $D/G$
\be S={1\over 2}I(f)-{1\over 4\pi}\int d\xi^+ d\xi^- \la 
\partial_+ f~f^{-1},R_-^a\ra (M_-^{-1})_{ab}\la f^{-1}\partial_- f,T^b\ra.\ee
The action of the dual $\sigma$-model on the coset $D/\ti G$ has 
the same form; just the generators $T^a$ of $\cg$ are replaced by the 
generators 
$\ti T_a$ of $\cgt$ and $f$ will parametrize $D/\ti G$ instead of $D/G$.

 In the special
case, when the decomposition $D=G\ti G=\ti GG$ holds globally (typically
$SL(n,C)$ doubles), the 
corresponding
cosets $D/G$ and $D/\ti G$ turn out to be the group manifolds $\ti G$ 
and $G$ respectively \cite{KS2} and the action (12) gives the standard 
pair of the  mutually dual lagrangians \cite{KS2}
\be \ti L={1\over 4\pi}({\cal R}^{-1}+\ti \Pi(\ti h))^{-1}(\partial_+ 
\ti h \ti h^{-1},
\partial_- \ti h \ti h^{-1})\ee
and  
\be L={1\over 4\pi}({\cal R}+\Pi(h))^{-1}(\partial_+ h h^{-1}, 
\partial_- h h^{-1}).\ee
Here $\ti h$ and $h$ respectively 
parametrize the $\ti G$ and $G$ group manifolds, 
$\ti\Pi(\ti h)$ and $\Pi(h)$ are the bivector fields
on the group manifold  $\ti G$ and $G$ which  respectively give
 the  standard Poisson-Lie brackets on $\ti G$ and $G$  \cite{D, KS2,KS3}
and  ${\cal R}$ is a bilinear form on $\cgt$ whose graph\footnote{Note that
${\cal R}(t,.)$ is an element of $\cg$.}   
$Span\{t+{\cal R}(t,.),t\in\cgt\}$ in 
$\cd$ coincides with the $+1$-eigenspace of $R$. 

It may seem that we have proved the canonical equivalence of the
$\sigma$-models (12) on the cosets $D/G$ and $D/\ti G$.
It is indeed true modulo one extremely important detail: the quantity
$\Lambda$ given by (10) has to fulfil a unit  monodromy constraint
\be P\exp{\int_{\gamma}\Lambda}=e\ee
where $e$ is the unit element of the group $G$, $P$ stands
for the path ordered exponential and $\gamma$ is a closed
path around the string world-sheet with a constant $\tau$ 
(the completely analoguos
statement is true also in the dual case $D/\ti G$). This constraint follows
from the obvious periodicity in $\sigma$ of the field $g(\tau,\sigma)$.
Thus we have established the duality between the classical
$\sigma$-models  with  the additional non-local constraints
imposed on their dynamics. What is the meaning of these constraints?

First of all we realize, that the equations of motions following
from the action (12) have the zero-curvature form (this property
was referred to as the `Poisson-Lie symmetry'' in \cite{KS2}):
\be d\lambda-\lambda^2=0.\ee
Here 
\be \lambda =\lambda_+ d\xi^+ +\lambda_- d\xi^-\ee
and
\be \lambda_{\pm}=-\la \partial_{\pm} f~f^{-1},R_{\mp}^a\ra
 (M_{\mp}^{-1})_{ab}T^b.\ee
Again, the completely analogous representation holds in the dual case.
It  follows that  the conjugacy class of the $G$-monodromy
\be P\exp{\int_{\gamma}\lambda}\ee
does not depend on the path $\gamma$ and it is therefore conserved in
time $\tau$. In particular, if the momonodromy is the unit element of $G$,
it is exactly conserved. Needless to say, the constraint  (10),(15) is just
the unit monodromy constraint of $\lambda$ for the particular path
of constant $\tau$. 

It may seem somewhat peculiar that we have established the classical duality
of two local $\sigma$-models only when certain non-local constraints 
are imposed on the dynamics. What it means classically that we
do not consider all possible motions of string which are allowed
 by the geometry
of the space-time? Would not it be too difficult
 to quantize such a constrained
theory? We believe that the answer to this question is surprisingly
simple: In many relevant cases it should be enough just to quantize
the unconstrained theory and the quantization itself would take care
for imposing the monodromy constraint! In order to clarify this
somewhat vague statement consider the well-known example of the
standard Abelian $\rho\to 1/\rho$ duality.

 3.  Consider the Abelian Drinfeld double $D_a$ which is just the
group $U(1)\times U(1)\equiv G\times \ti G$. The group manifold is 
topologically the ordinary
torus and we choose  its explicit parametrization as
\be l=e^{iaT}e^{i\ti a\ti T},\ee
where $T$ and $\ti T$ are the algebra generators of $\cg$ and $\cgt$ 
respectively and   $a,\ti a\in [0,2\pi)$.
The bilinear form in $\cd$ we define as
\be \la T,T\ra=\la \ti T,\ti T\ra=0, \quad \la T,\ti T\ra =1.\ee
We choose the normalized vectors $R_{\pm}\in \cd$ as
\be R_{\pm}=\sqrt{\rho\over 2}T\pm{1\over \sqrt{2 \rho}}\ti T,\ee
the parametrization of $D/\ti G$ clearly to be $f=e^{iaT},  a\in[0,2\pi)$  
(and analogously for $D/G$) and work out directly the mutually dual 
models (12):
\be S=-{1\over 4\pi\rho}\int d\xi^+d\xi^- \partial_+a\partial_-a,\ee
\be \ti S=-{\rho\over 4\pi}\int d\xi^+d\xi^- 
\partial_+\ti a\partial_-\ti a.\ee
The corresponding connections $\lambda_{\pm}$ and $\ti\lambda_{\pm}$
read
\be \ti\lambda_{\pm}=\pm {i\over \rho}\partial_{\pm}a\ti T,\ee
\be \lambda_{\pm}=\pm i\rho\partial_{\pm}\ti a T.\ee

It is now  easy to quantize the free field theory (23) (the case 
 (24) differs just by the change $\rho\to 1/\rho$). Consider the
mode expansion of the field $a$
\be a=a_0+p_L \xi^- +p_R\xi^+  +osc_L +osc_R; \ee
the exact form of the oscillator terms is irrelevant for our purposes.
The single-valuedness of the field $a(\sigma,\tau)$ requires that the winding
number ${1\over 2}(p_L-p_R)$ be integer;
however, classically there is no constraint on the momentum ${1\over 2}
(p_L+p_R)$. In the quantum case, however,  the spectrum
of the momentum read:
\be {1\over 2}(p_L+p_R)= m\rho , \quad m\in {\bf Z}\ee
But then from (25) and (27) we get for the monodromy 
$$P\exp{\oint \ti\lambda}=\ti e,$$
where $\ti e$ is  the unit element of the group $\ti G$ (or of $G$
in the dual picture). We witness that there is no necessity of imposing
the constraint of the unit monodromy at the quantum level, because
the process of the quantization itself takes care of it.

4. In our previous works \cite{KS2,KS3}, we have referred to
the monodromy $P\exp{\oint\lambda}$ as to 
the non-commutative $G$-valued momentum
of the string. It turned out that the geometry of the targets of the
dualizable $\sigma$-models allows to write the equations of motions
as the zero-curvature condition for the connection $\lambda$, which in 
turn means that the conjugacy
class of the
monodromy of $\lambda$
is the conserved quantity - the non-commutative momentum. 
In the just-described
Abelian case, the zero-curvature condition coincides with the $U(1)$-current
conservation equation which implies the conservation of the total
momentum\footnote{Note a terminological difficulty: the `non-commutative
momentum' becomes in the $U(1)$ case just the exponent of the $2\pi i$
times the standard momentum.} $p$ and hence of the monodromy 
$\exp{2\pi ip T}$. 
In the Abelian quantum theory the monodromy is always $e$ but we also
observe that this does not mean that there is a single momentum state!
In fact, it appears to be more natural to evaluate the monodromy in 
the universal
covering group of $G$ and to understand the non-commutative momentum as 
an element of this cover. The constraint of the unit
$G$-valued monodromy then means that only those cover-valued monodromies 
which get projected
to the unit element of $G$ are allowed.

It is easy to  interpret  the different classical 
(Abelian) momentum  states, which
give the same unit $G$-monodromy. 
  The state with the Abelian momentum $n/\rho$ (cf. model (24)) correspond
to a loop which wraps $n$-times along the homology cycle in the double
generated by $T$ (record that the winding states of the model (24) wrap
along the cycle $\ti T$). Thus we observe the perfect duality in the
classical phase space: from the point of view of the model (24) the
momentum and the winding states correspond to the homology cycles
$T$ and $\ti T$ respectively, whileas for the dual model (23) the role of the
homology cycles gets interchanged. 

Consider now the  fundamental group $\Gamma(D)$ of any Drinfeld double 
(in the Abelian case it 
is just $\bf Z\oplus \bf Z$). The phase space of the
model (2) or (12)
  decomposes into disconnected sectors labelled by the elements
of $\Gamma(D)$. Upon taking
the coset $D/G$, some of the loops from $\Gamma(D)$ will remain incontractible
in $D/G$ and we naturally interpret them as the winding states in $D/G$ target.
On the other hand, those loops from $\Gamma(D)$ which become contractible
after projection to the coset $D/G$ we interpret as the momentum states.
Clearly, this interpretation is not duality invariant. For instance, if we 
consider
the dual coset $D/\ti G$ in the Abelian case then the role of the momentum
and the winding states gets precisely interchanged. This is the famous
phenomenon of the Abelian momentum-winding exchange \cite{SS}. 

In the non-Abelian
case the situation is very similar though few other scenarios may appear
than just the complete interchange of the momentum and winding states. 
In what follows we shall consider only the case when all involved groups
$D$,$G$ and $\ti G$ are connected but not necessarily simply connected.
The momentum-winding interpretation is then governed by the 
following long exact 
homotopy sequence \cite{Schw}:
\be \pi_2(D)=0\to \pi_2(D/G)\to\pi_1(G)\to\pi_1(D)\to\pi_1(D/G)\to 0=
\pi_0(G).\ee
Note  that $\pi_2$ of any Lie
group vanishes.
We can rewrite this sequence
as follows
\be 0\to \pi_1(G)/\pi_2(D/G)\to \pi_1(D)\to \pi_1(D/G)\to 0\ee
and observe that the (Abelian) group $\pi_1(D)$ is an extension of the 
(Abelian) group $\pi_1(D/G)$ by the (Abelian) group $\pi_1(G)/\pi_2(D/G)$.
Note the role of $\pi_2(D/G)$: Its possible non-vanishing means that
some non-contractible cycles in $G$ can be contracted in $D$ upon embedding
of $G$ in $D$. Thus we can conclude that the winding modes of 
the $\sigma$-model on $D/G$ are the elements of $\pi_1(D/G)$ and the
momentum modes are the elements of $\pi_1(G)/\pi_2(D/G)$. In the dual case 
 the winding (momentum) modes are the elements
of $\pi_1(D/\ti G)$ ($\pi_1(\ti G)/\pi_2(D/\ti G)$). Thus we
 observe that  the partition of $\pi_1(D)$ into the momentum and winding
 piece depends on the target.
  There may be an element of $\pi_1(D)$ which is a momentum mode from
the point of view of $D/G$ but a winding mode for $D/\ti G$. 

In the case of the 
 traditional non-Abelian duality \cite{OQ,GR} there was  a little room
to discover the momentum-winding exchange, because the Drinfeld
double is then  the  cotangent bundle of some compact
group $G$ and the role of $\ti G$ is played by its  coalgebra $\cg^*$
viewed as the commutative group. The point is that $\pi_1(D)$
is usually quite poor in this case. For instance, for a simply connected $G$
(and hence $D$) 
 there is no trace of the non-trivial
momentum or winding states whatsoever.

5. Example $D=SL(2,R)\times SL(2,R)$.
Consider the Lie algebra $sl(2,R)$  defined by
\be [H,E_{\pm}]=\pm 2E_{\pm}, \qquad [E_+,E_-]=H, \ee
and equipped with  the standard Killing-Cartan non-degenerate symmetric
invariant
bilinear form
\be \la E_+,E_-\ra=1, \qquad \la H,H\ra=2.\ee
 The direct sum of the two copies of $sl(2,R)$
 \be \cd =sl(2,R)\oplus sl(2,R)\ee
with the bilinear form (also denoted by $\la .,.\ra$)
\be \la (x_1,x_2),(y_1,y_2)\ra=\la x_1,y_1\ra -\la x_2,y_2\ra\ee
is the Lie algebra of the  Drinfeld double $D$.
The notation $(x_1,x_2)\in\cd$ obviously means that $x_1$ ($x_2$) is from
the first (second) copy of $sl(2,R)$ in  (33).
The decomposition of the double into the pair of the maximally
 isotropic subalgebras
is given by
\be \cd =sl(2,R)_{ diag}  +  b_2\ee
where $sl(2,R)_{ diag}$  is generated by
\be \ti T_0 ={1\over 2}(H,H),\quad \ti T_+= (E_+,E_+), \quad \ti T_-=
 (E_-,E_-)\ee
and $b_2$ (which is the Lie algebra of the Borel group $B_2$ consisting
of upper-triangular $2\times 2$ complex matrices with real positive 
diagonal elements and unit determinant)  by
\be T^0={1\over 2}(H,-H), \quad T^+ = (0,-E_-), \quad  T^- =(E_+,0).\ee
 
The homotopy groups of $D$, $SL(2,R)$, $B_2$ and of the relevant
cosets $D/B_2$, $D/SL(2,R)_{diag}=SL(2,R)$ read
$$\pi_1(D)=\pi_1(D/B_2)={\bf Z}\oplus {\bf Z}, \quad \pi_1(SL(2,R))=
\pi_1(D/SL(2,R)_{diag})=
{\bf Z},$$
\be  \quad \pi_1(B_2)=\pi_2(D/B_2)=\pi_2(D/SL(2,R)_{diag})=0.\ee
We conclude that the $\sigma$-model (12) on $D/B_2$ has two
types of winding states and on $D/SL(2,R)_{diag}$ \footnote{For a particular 
choice of $R_{\pm}$ the model on $D/Sl(2,R)_{diag}$
 is just the standard WZNW model on $SL(2,R)$
 \cite{AKT}.} just one type.
Under duality, the winding states of $D/B_2$ of one type correspond to the
non-commutative momentum states of $D/SL(2,R)_{diag}$ and the winding 
states of the other type remain the winding states on $D/SL(2,R)_{diag}$.

 6. Concluding remarks:

\noindent i) In the previous discussion we have been always talking about the
Drinfeld double. However, the careful reader might have remarked that
the described construction of the dual $\sigma$-models can be repeated
in a more general setting. Essentially, we just require that $D$
is  a $2n$-dimensional 
Lie group with a non-degenerate symmetric $ad$-invariant bilinear
form
on its algebra which in turn admits at least two 
different $n$-dimensional isotropic subalgebras.
Clearly, the Drinfeld double is always an example of such a situation,
however, there may be more examples of this type. For instance, consider
a compact connected simple Lie group $G$ and put $D=G\times G$
with the  bilinear form on $\cd=\cg\oplus\cg$ given just by the difference
of the standard Killing-Cartan forms  on the first and second copy
of $\cg$ respectively (like in the $SL(2,R)\times SL(2,R)$-case considered
previously). 
We may consider two different embeddings of the group $G$ in $D$.
First one is the standard diagonal embedding and in the second,
$G$ is identically embedded into the first copy of $G$ and up to twist by
an outer automorphism into the second copy of $G$. It is obvious that 
both embeddings are  isotropic at the level of Lie algebra
$\cd =\cg\oplus\cg$. This construction will most probably be connected with 
the Kiritsis-Obers duality \cite{KO}.
 
\noindent ii) We hope to supply
a detailed quantum picture of the presented structures in a near future.
 The most obvious open problem is  the
quantum status of the unit monodromy constraint. After previous experience
with the naturalness of the structure of PL $T$-duality, we expect   this
issue will be settled   and      the  PL $T$-duality  will find
interesting applications in both quantum field and string theories. 

\vskip1pc
 We thank A. Connes and E. Kiritsis for discussions.

\end{document}